\documentclass[aps,pre,reprint,showpacs,groupedaddress,superscriptaddress,floatfix]{revtex4-1}
\usepackage[dvipdfmx]{graphicx}
\usepackage{amsmath,amssymb}
\usepackage{bm}
\usepackage{url}
\usepackage{booktabs}
\usepackage{braket}
\usepackage{color, ulem}

\begin{document}
%
%
%
%
\title{Rattleback dynamics and its reversal time of rotation}
%
%
\author{Yoichiro \surname{Kondo}}
\email[]{ykondo@stat.phys.kyushu-u.ac.jp}
\author{Hiizu \surname{Nakanishi}}
\affiliation{Department of Physics, Kyushu University, 33, Fukuoka 819-0395, Japan}
%
%
\date{\today}
\begin{abstract}
A rattleback is a rigid, semi-elliptic toy which exhibits unintuitive
behavior; when it is spun in one direction, it soon begins pitching and
stops spinning, then it starts to spin in the opposite direction, but in
the other direction, it seems to spin just steadily. This puzzling
behavior results from the slight misalignment between the principal
axes for the inertia and those for the curvature;
the misalignment couples the spinning 
with the pitching and the rolling oscillations. It has been shown
that under the no-slip condition and without dissipation the spin can
reverse in both directions, and Garcia and Hubbard obtained the formula
for the time required for the spin reversal $t_r$ [Proc. R. Soc. Lond. A
\textbf{418}, 165 (1988)].
In this work, we reformulate the rattleback dynamics in a
physically transparent way
and reduce it to a three-variable dynamics  for spinning,
pitching, and rolling.  
We obtain an expression of the Garcia-Hubbard formula for $t_r$ by a simple
product of four factors: (1) the misalignment angle, (2) the difference
in the inverses of inertia moment for the two oscillations, (3)
that in the radii for the two principal curvatures, and (4) the squared
frequency of the oscillation.
We perform extensive numerical simulations to examine
validity and limitation of the formula, and find that (1) the Garcia-Hubbard
formula is good for both spinning directions in the small spin and 
small oscillation regime, but (2) in the fast spin regime 
especially for the steady
direction, the rattleback may not reverse and shows a rich variety of
dynamics including steady spinning, spin wobbling, and chaotic behavior
reminiscent of chaos in a dissipative system.

\end{abstract}

\maketitle
%
%
%
\section{Introduction}
\label{sec:introduction}
Spinning motions of rigid bodies have been studied for centuries and
still are drawing interest in recent years, including the motions of
Euler's disks \cite{Moffatt2000}, spinning eggs \cite{Moffatt2002}, and
rolling rings \cite{Jalali2015}, to mention just a few. Also,
macroscopic systems which convert vibrations to rotations have been
studied in various context such as a circular granular ratchet
\cite{Heckel2012}, and bouncing dumbbells, which show a cascade of
bifurcations \cite{Kubo2015}. Another interesting example of rigid body
dynamics which involves such oscillation-rotation coupling is a
rattleback, also called as a celt or wobble stone, which is a
semi-elliptic spinning toy [Fig.~\ref{fig:notation}(a)]. It spins
smoothly when spun in one direction; however, when spun in the other
direction, it soon starts wobbling or rattling about its short axis and
stops spinning, then it starts to rotate in the opposite direction. One
who has studied classical mechanics must be amazed by this reversal in
spinning, because it apparently seems to violate the angular momentum
conservation, and the chirality emerges from a seemingly symmetrical
object.

There are three requirements for a rattleback to show this reversal of
rotation: (1) the two principal curvatures of the lower surface should
be different, (2) the two horizontal principal moments of inertia
should also be different, and (3) the principal axes of inertia should
be misaligned to the principal directions of curvature. These
characteristics induce the coupling between the spinning motion and the
two oscillations: the pitching about the short horizontal axis and the
rolling about the long horizontal axis.  The coupling is asymmetric,
i.e., the oscillations cause torque around the spin axis and the
signs of the torque are opposite to each other.  This also means
that either the pitching or the rolling is excited depending on the
direction of the spinning. We will see that the spinning couples with
the pitching much stronger than that with the rolling; therefore, it
takes much longer time for spin reversal in one direction than in the
other direction, and that is why most rattlebacks reverse only for one way
before they stop by dissipation.

In the 1890s, a meteorologist, Walker, performed the first quantitative
analysis of the rattleback motion \cite{Walker1896}. Under the
assumptions that the rattleback does not slip at the contact point and
that the rate of spinning speed changes much slower than other time
scales, he linearized the equations of motion and showed that
either the pitching or the rolling becomes unstable depending on the
direction of the spin. More detailed analyses were performed by Bondi
\cite{Bondi1986}, and recently by Wakasugi \cite{WakasugiH23}. Case and
Jalal \cite{Case2014} derived the growth rate of instability at slow
spinning. Markeev \cite{Markeev1984}, Pascal \cite{Pascal1983}, and
Blackowiak et al. \cite{Blackowiak1997} obtained the equations of
the spin motion by extracting the slowly varying amplitudes of the fast
oscillations of the pitching and the rolling. Moffatt and Tokieda
\cite{MoffattTokieda2008} derived similar equations to those of Markeev
\cite{Markeev1984} and Pascal \cite{Pascal1983}, and pointed out the
analogy to the dynamo theory. Garcia and Hubbard
\cite{GarciaHubbard1988} obtained the expressions of the averaged
torques generated by the pure pitching and the rolling, and derived the
formula for spin reversal time.
 
As the first numerical study, Kane and Levinson \cite{Kane1982}
simulated the energy-conserving equations and showed that the rattleback
changes its spinning direction indefinitely for certain parameter values
and initial conditions. They also demonstrated the coupling between the
oscillations and the spinning by showing that it starts to rotate when
it begins with pure pitching or rolling, but the direction of the
rotation is different between pitching and rolling. Similar simulations
were performed by Lindberg and Longman independently
\cite{Lindberg1983}. Nanda \textit{et al.} simulated the spin resonance of the
rattleback on a vibrating base \cite{Nanda2016}.

Energy-conserving dynamical systems usually conserve the phase volume,
but the present rattleback dynamics does not explore the whole phase
volume with a given energy because of a non-holonomic constraint due to
the no-slip condition. Therefore, the Liouville theorem does not hold,
and such a system has been shown to behave much like dissipative
systems. Borisov and Mamaev in fact reported the existence of ``strange
attractor'' for certain parameter values in the present system
\cite{Borisov2003}. The no-slip rattleback system has been actively
studied in the context of chaotic dynamics during the last decade
\cite{Borisov2006,Borisov2014}.

Effects of dissipation at the contact point have been investigated in
several works. Magnus \cite{Magnus1974} and Karapetyan
\cite{Karapetyan1981} incorporated a viscous type of friction force
proportional to the velocity. Takano \cite{Takano2014} determined the
conditions under which the reversal of rotation occurs with the viscous
dissipation. Garcia and Hubbard \cite{GarciaHubbard1988} simulated
equations with aerodynamic force, Coulomb friction in the spinning, and
dissipation due to slippage, then they compared the results with a real
rattleback. The dissipative rattleback models based on the contact
mechanics with Coulomb friction have been developed by
Zhuravlev and Klimov \cite{Zhuravlev2008} and Kudra and Awrejcewicz
\cite{Awrejcewicz2012, Kudra2013, Kudra2015}.

This paper is organized as follows. In the next section, we
reformulate the rattleback dynamics under the no-slip and no dissipation
condition in a physically transparent way. In the small-spin and
small-oscillation approximation, the dynamics is reduced to a
simplified three-variable dynamics. We then focus on the time required
for reversal, or what we call \textit{the time for reversal},
which is the most evident quantity that characterizes
rattlebacks, and obtain a concise expression for the Garcia-Hubbard
formula for the time for reversal \cite{GarciaHubbard1988}. In
Sec.~\ref{sec:simulation}, the results of the extensive numerical
simulations are presented for various model parameters and initial
conditions in order to examine the validity and the limitation of the
theory.  Discussions and conclusion are given in
Sec.~\ref{sec:discussion} and Sec.~\ref{sec:conclusion}, respectively.
%
\section{Theory}
\label{sec:theory}
\subsection{Equations of motion}
%
\begin{figure}
 \includegraphics[width=7cm]{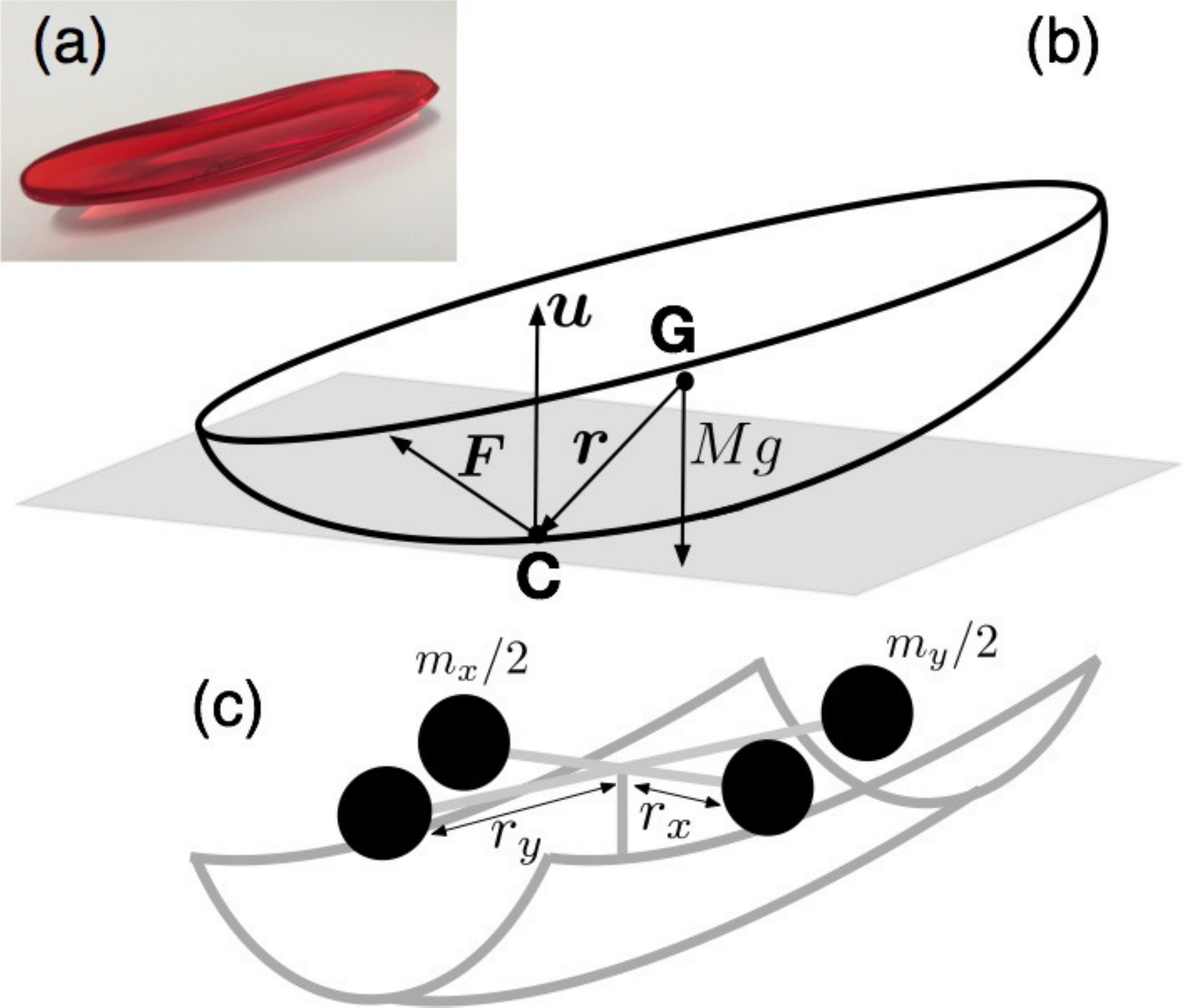}
 \caption{\label{fig:notation}(a) A commercially available rattleback made of plastic. (b) Notations of the rattleback. (c) A schematic illustration of the shell-dumbbell model.}
\end{figure}
%
We consider a rattleback as a rigid body, whose configuration can be
represented by the position of the center of mass G and the Euler
angles; both of them are obtained by integrating the velocity of the
center of mass $\bm{v}$ and the angular velocity $\bm{\omega}$ around it
\cite{Goldstein2002}.

We investigate the rattleback motion on a horizontal plane, assuming
that it is always in contact with the plane at a single point C without
slipping. We ignore dissipation, then all the forces that act on
the rattleback are the contact force $\bm{F}$ exerted by the plane
at C and the gravitational force $-Mg\bm{u}$, where $\bm{u}$ represents
the unit vertical vector pointing upward
[Fig.~\ref{fig:notation}(b)]. Therefore, the equations of motion are
given by
\begin{align}
 \frac{d(M\bm{v})}{dt} &= \bm{F} - Mg\bm{u},\label{eq:em-1}\\
 \frac{d(\hat{I}\bm{\omega})}{dt} &= \bm{r} \times \bm{F}, \label{eq:em-2}
\end{align}
where $M$ and $\hat{I}$ are the mass and the inertia tensor around G,
respectively, and $\bm{r}$ is the vector from G to the contact point C.

\begin{figure}
\includegraphics[width=8cm]{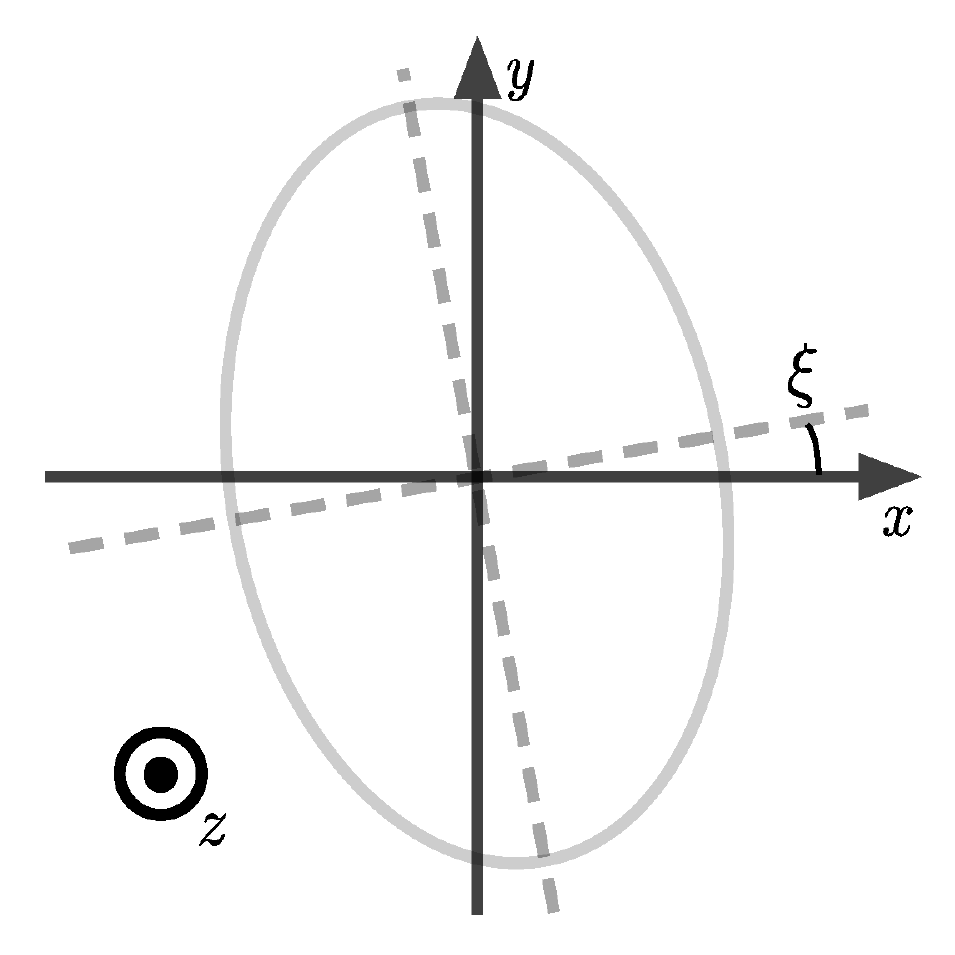}
 \caption{(color online) \label{fig:coordinate}A body-fixed co-ordinate viewed from below. The dashed lines indicate the principal directions of curvature, rotated by $\xi$ from the principal axes of inertia (the $x$-$y$ axes).}
\end{figure}

The contact force $\bm{F}$ is determined by the conditions of the
contact point; our assumptions are that (1) the rattleback is always in
contact at a point with the plane, and (2) there is no slip at the
contact point. The second constraint is represented by the relation
%
\begin{equation}
 \bm{v} =  \bm{r} \times \bm{\omega}.\label{eq:no-slip}
\end{equation}
Before formulating the constraint (1), we specify the co-ordinate
system.  We employ the body-fixed co-ordinate with the origin being the
center of mass G, and the axes being the principal axes of inertia; the
$z$ axis is the one close to the spinning axis pointing downward, and
the $x$ and $y$ axes are taken to be $I_{xx} > I_{yy}$
(Fig. \ref{fig:coordinate}).

In this co-ordinate, the lower surface function of the rattleback is assumed to be given by 
\begin{equation}
 f(x,y,z) = 0,\label{eq:def-z}
\end{equation}
where
\begin{equation}
	f(x,y,z) \equiv \frac{z}{a} - 1 + \frac{1}{2a^2}(x,\,y)\hat{R}(\xi)\hat{\Theta}\hat{R}^{-1}(\xi)\begin{pmatrix}x\\y\\\end{pmatrix},
\end{equation}
with
\begin{align}
\hat{R}(\xi) \equiv
\begin{pmatrix}
\cos\xi, & -\sin\xi \\
\sin\xi, & \cos\xi \\
\end{pmatrix}, \quad 
\hat{\Theta} \equiv
\begin{pmatrix}
\theta, & 0 \\
0 & \phi \\
\end{pmatrix}.
\end{align}
Here $a$ is  the distance between G and the surface at $x=y=0$, and $\xi$ is the \textit{skew angle} by which the principal directions of curvature are rotated from the $x$-$y$ axes, which we choose as the principal axes of inertia (Fig. \ref{fig:coordinate}). $\theta/a$ and $\phi/a$ are the principal curvatures at the bottom, namely at $(0,0,a)^{t}$. 

Now, we can formulate the contact point condition (1); the components of
the contact point vector $\bm{r}$ should satisfy Eq.~(\ref{eq:def-z}),
and the normal vector of the surface at C should be parallel to the
vertical vector $\bm{u}$. Thus we have
\begin{equation}
 \bm{u} \parallel \nabla f, 
\end{equation}
which gives the relation
\begin{equation}
	\frac{\bm{r}_{\perp}}{a} = \frac{1}{u_{z}} \hat{R}(\xi)\hat{\Theta}^{-1}\hat{R}^{-1}(\xi)\bm{u}_{\perp} \label{eq:def-xy},
\end{equation}
where $\bm{a}_{\perp}$ represents the $x$ and $y$ components of a
vector $\bm{a}$ in the body-fixed co-ordinate.

Before we proceed, we introduce a dotted derivative of a vector $\bm{a}$
defined as the time derivative of the vector components in the
body-fixed co-ordinate. This is related to the time derivative by
\begin{equation}
 \frac{d\bm{a}}{dt} = \dot{\bm{a}} + \bm{\omega} \times \bm{a}.
\end{equation}
Note that the vertical vector $\bm{u}$ does not depend on time, thus we have
\begin{equation}
 \frac{d\bm{u}}{dt} = \dot{\bm{u}} + \bm{\omega} \times \bm{u} = \bm{0}. \label{eq:diff-u} 
\end{equation}
These conditions, i.e., the no-slip condition (\ref{eq:no-slip}), the conditions of the contact point (\ref{eq:def-z}) and (\ref{eq:def-xy}), and the vertical vector condition (\ref{eq:diff-u}), close the equations of motion (\ref{eq:em-1}) and (\ref{eq:em-2}).

Following Garcia and Hubbard \cite{GarciaHubbard1988}, we describe the rattleback dynamics by $\bm{u}$ and $\bm{\omega}$. 
The evolution of $\bm{\omega}$ is obtained as
\begin{multline}
 \hat{I} \dot{\bm{\omega}} -  M\bm{r} \times (\bm{r} \times \dot{\bm{\omega}})
 =  -  \bm{\omega} \times (\hat{I}\bm{\omega}) \\
 + M\bm{r}\times(\dot{\bm{r}}\times \bm{\omega} + \bm{\omega}\times (\bm{r} \times \bm{\omega})) + Mg\bm{r}\times\bm{u} \label{eq:diff-omega}
\end{multline}
by eliminating the contact force $\bm{F}$ from the equations of motion (\ref{eq:em-1}) and (\ref{eq:em-2}), and using the no-slip condition (\ref{eq:no-slip}).
The state variables $\bm{u}$ and $\bm{\omega}$ can be determined by Eqs.~(\ref{eq:diff-u}) and (\ref{eq:diff-omega}) with the contact point conditions (\ref{eq:def-z}) and (\ref{eq:def-xy}).
%

The rattleback is characterized by the inertial parameters $M$,
$I_{xx}$, $I_{yy}$, $I_{zz}$, the geometrical parameters $\theta$,
$\phi$, $a$, and the skew angle $\xi$.  For the stability of the
rattleback, both of the dimensionless curvatures $\theta$ and
$\phi$ should be smaller than $1$; without loss of generality, we assume
\begin{equation}
 0 < \phi < \theta < 1,
\end{equation}
then, it is enough to consider 
\begin{equation}
 -\frac{\pi}{2} < \xi < 0,
\end{equation}
for the range of the skew angle $\xi$. The positive $\xi$ case can be obtained by the reflection with respect to the $x$-$z$ plane.

At this stage, we introduce the dimensionless inertial parameters
$\alpha$, $\beta$, and $\gamma$ for later use after Bondi
\cite{Bondi1986} as
\begin{equation}
\alpha \equiv \frac{I_{xx}}{Ma^{2}}+1, \ \beta \equiv \frac{I_{yy}}{Ma^{2}}+1, \ \gamma \equiv \frac{I_{zz}}{Ma^2},\label{eq:def-abg}
\end{equation}
which are dimensionless inertial moments around the contact
point C.  Note that
\begin{equation}
\alpha > \beta > 1,
\end{equation}
because we have assumed $I_{xx} > I_{yy}$. 

\subsection{Small amplitude approximation of oscillations under $\omega_{z}=0$}
\label{subsec:linearization}
We consider the oscillation modes in the case of no spinning
$\omega_{z} = 0$ in the small amplitude approximation, namely, in the
linear approximation in $|\omega_{x}|,\,|\omega_{y}|\ll\sqrt{g/a}$,
which leads to $|x|,\,|y| \ll a$, $|u_{x}|,\, |u_{y}| \ll 1 $, and
$u_{z} \approx -1$.
In this regime, the $x$ and $y$ components of Eq.~(\ref{eq:diff-u}) can
be linearized as
\begin{equation}
 \dot{\bm{u}}_{\perp} \approx \hat{\varepsilon}\,\bm{\omega}_{\perp}, \quad
\hat{\varepsilon} \equiv
\begin{pmatrix}
0, & 1 \\
-1, & 0 \\
\end{pmatrix} = \hat{R}(-\pi/2).\label{eq:lin-u}
\end{equation}
By using Eq.~(\ref{eq:def-xy}) with $u_{z} \approx -1$,  Eq.~(\ref{eq:diff-omega}) can be linearized as
\begin{align}
 \hat{J}\,\dot{\bm{\omega}}_{\perp} &\approx \frac{g}{a^2} (\bm{r}\times\bm{u})_{\perp}\notag\\
 &= -\frac{g}{a}\hat{\varepsilon}\,[-\hat{R}(\xi)\hat{\Theta}^{-1}\hat{R}^{-1}(\xi) +1 ] \bm{u}_{\perp},\label{eq:lin-omg1}
\end{align}
with the inertial matrix
\begin{equation}
\hat{J} \equiv 
\begin{pmatrix}
\alpha, & 0 \\
0, & \beta \\
\end{pmatrix}.
\end{equation}

From the linearized equations (\ref{eq:lin-u}) and (\ref{eq:lin-omg1}), we obtain 
\begin{equation}
	\hat{J}\ddot{\bm{\omega}}_{\perp}= - \frac{g}{a}(\hat{\Gamma} -1) \bm{\omega}_{\perp},\label{eq:lin-omg2}
\end{equation}
where
\begin{equation}
\hat{\Gamma} \equiv \hat{R}(\xi +\pi/2) \hat{\Theta}^{-1}\hat{R}^{-1}(\xi+\pi/2).
\end{equation}
At this point, it is convenient to introduce the bra-ket notation for the row and column vector of $\bm{\omega}_{\perp}$ as $\bra{\omega_{\perp}}$ and $\ket{\omega_{\perp}}$, respectively. With this notation, Eq.~(\ref{eq:lin-omg2}) can be put in the form of 
\begin{align}
    \ket{\ddot{\tilde{\omega}}_{\perp}}= -\hat{H}\ket{\tilde{\omega}_{\perp}},
\label{eq:lin-omg3}
\end{align}
with
\begin{align}
   \ket{\tilde{\omega}_{\perp}} \equiv \hat{J}^{1/2}\ket{\omega_{\perp}},
   \quad \hat{H} \equiv \frac{g}{a}\hat{J}^{-1/2}(\hat{\Gamma} -1 )\hat{J}^{-1/2},
\label{eq:lin-omg4}
\end{align}
where $\hat{H}$ is symmetric. The eigenvalue equation
\begin{align}
 \hat{H} \ket{\tilde{\omega}_{j}}= \omega_{j}^2 \ket{\tilde{\omega}_{j}}
\label{eq:def-omgpr}
\end{align}
determines the two oscillation modes with $j=p$ or $r$, whose frequencies are given by
\begin{equation}
\omega_{p,r}^2 = \frac{1}{2}\left[(H_{11}+H_{22})
      \pm \sqrt{(H_{11}-H_{22})^2 + 4H_{12}^2}\right]
\label{eq:def-omega_pr}
\end{equation}
with
\begin{equation}
\omega_{p} \ge \omega_{r}.
\label{ineq:omega_pr}
\end{equation}
Here, $H_{ij}$ denotes the $ij$ component of $\hat H$.
The orthogonal condition for the eigenvectors
$\ket{\tilde{\omega}_{p}}$ and $\ket{\tilde{\omega}_{r}}$ can be
written using $\hat{\varepsilon}$ as
\begin{align}
 \ket{\tilde{\omega}_{p}} &= \hat{\varepsilon} \ket{\tilde{\omega}_{r}},\quad
 \ket{\tilde{\omega}_{r}} = -\hat{\varepsilon} \ket{\tilde{\omega}_{p}},  \\
 \bra{\tilde{\omega}_{r}} &=  \bra{\tilde{\omega}_{p}}\hat{\varepsilon}, \quad
 \bra{\tilde{\omega}_{p}} =  -\bra{\tilde{\omega}_{r}}\hat{\varepsilon}. 
\end{align}

In the case of zero skew angle, $\xi=0$, we have
\begin{align}
\omega_p^2 &=\left({g\over a}\right){1/\phi-1\over\alpha}\equiv \omega_{p0}^2,
\label{def:omega_p0}
\\
\omega_r^2 &=\left({g\over a}\right){1/\theta-1\over\beta}\equiv \omega_{r0}^2,
\label{def:omega_r0}
\end{align}
and the eigenvectors $\ket{\omega_{p}}$ and $\ket{\omega_{r}}$ are
parallel to the $x$ and the $y$ axes, thus these modes correspond to the
pitching and the rolling oscillations, respectively.  This correspondence
holds for $|\xi|\ll 1$ and $\omega_{p0}>\omega_{r0}$ as for a typical
rattleback parameter, the case we will discuss mainly in the following
\footnote{Note that in the atypical case of
$\omega_{p0}<\omega_{r0}$, i.e.  the pitching is slower than
the rolling, we have $\omega_p\approx\omega_{r0}$ and
$\omega_r\approx\omega_{p0}$ for $|\xi|\ll 1$ because $\omega_p>\omega_r$
by Eq.~(\ref{ineq:omega_pr}).}.
%
\subsection{Garcia and Hubbard's theory for the time for reversal}
\label{subsec:GHtheory}
Based on our formalism, it is quite straightforward to derive Garcia and Hubbard's formula for the reversal time of rotation.
%
\subsubsection{Asymmetric torque coefficients}
Due to the skewness, the pitching and the rolling are coupled with the
spinning motion. We examine this coupling in the case of
$\omega_{z} = 0$ by estimating the averaged torques around the vertical
axis caused by the pitching and the rolling oscillations. From
Eqs.~(\ref{eq:em-1}) and (\ref{eq:em-2}) and the no-slip condition
Eq.~(\ref{eq:no-slip}), the torque around $\bm{u}$ is given by
\begin{align}
 T &\equiv \bm{u}\cdot(\bm{r} \times \bm{F})
 \approx - Ma^2 [\dot{\bm{\omega}}_{\perp}\cdot\hat{\varepsilon}(\hat{\Gamma} - 1)\hat{\varepsilon} \,\bm{u}_{\perp}\,],\label{eq:ave-torque}
\end{align}
within the linear approximation in $\bm{\omega}_{\perp}$, $\bm{u}_{\perp}$, and $\bm{r}_{\perp}$ discussed in Sec.~\ref{subsec:linearization}.

We define the \textit{asymmetric torque coefficients} $K_{p} $ and $K_{r}$ for each mode by 
\begin{equation}
  -K_{p} \equiv \frac{\overline{T}_{p}}{\overline{E}_{p}}, \qquad 
  K_{r} \equiv \frac{\overline{T}_{r}}{\overline{E}_{r}},
  \label{eq:def-kpr}
\end{equation} 
where $\overline{T}_{j}\ (j=p\text{~or~}r)$ is the averaged torque over the oscillation period  generated by each mode, and $\overline{E}_{j}$ is the corresponding averaged oscillation energy which can be estimated within the linear approximation as 
\begin{align}
 \overline{E}  &\approx Ma^2 (\alpha \overline{\omega_{x}^2} + \beta \overline{\omega_{y}^2} ). \label{eq:ave-ene}
\end{align}
The minus sign for the definition of $K_{p}$ is inserted in order that both $K_{p}$ and $K_{r}$ should be positive for typical rattleback parameters as can be seen below. Note that the asymmetric torque coefficients are dimensionless.

From Eqs.~(\ref{eq:ave-torque}) and (\ref{eq:ave-ene}), $-K_{p}$ is given by
\begin{align}
-K_{p} &= \frac{\braket{\omega_{p}|\,\hat{\varepsilon}(\hat{\Gamma} - 1)\hat{\varepsilon}\hat{\varepsilon}\,|\omega_{p}}}{\braket{ \omega_{p}|\hat{J}|\omega_{p}}}\notag\\
&= -\frac{(a/g)\braket{\tilde{\omega}_{p}|\,\hat{J}^{-1/2}\hat{\varepsilon}\hat{J}^{1/2}\hat{H}\,|\tilde{\omega}_{p}}}{\braket{\tilde{\omega}_{p}|\tilde{\omega}_{p}}}\label{eq:kp-1}\\
&= - \omega_{p}^2 \,\frac{(a/g)\braket{\tilde{\omega}_{p}|\,\hat{J}^{-1/2}\hat{\varepsilon}\hat{J}^{1/2}\,|\tilde{\omega}_{p}}}{\braket{\tilde{\omega}_{p}|\tilde{\omega}_{p}}}.
\label{eq:kp-2}
\end{align}
In the same way, $K_{r}$ is given by
\begin{align}
K_{r} &= -\frac{(a/g)\braket{\tilde{\omega}_{r}|\hat{J}^{-1/2}\hat{\varepsilon}\hat{J}^{1/2}\hat{H}|\tilde{\omega}_{r}}}{\braket{\tilde{\omega}_{r}|\tilde{\omega}_{r}}}\label{eq:kr-1}\\
&= \omega_{r}^2 \,\frac{(a/g)\braket{\tilde{\omega}_{p}|(\hat{J}^{-1/2}\hat{\varepsilon}\hat{J}^{1/2})^{\dagger}|\tilde{\omega}_{p}}}{\braket{\tilde{\omega}_{p}|\tilde{\omega}_{p}}}.
\label{eq:kr-2}
\end{align}
Equations (\ref{eq:kp-1})--(\ref{eq:kr-2}) yield simple relations for $K_{p}$ and $K_{r}$ as
\begin{align}
\frac{K_{p}}{K_{r}} =  \frac{\omega_{p}^2}{\omega_{r}^{2}} \label{eq:k-rat}
\end{align}
and
\begin{align}
 K_{p} - K_{r}&= \frac{(a/g)}{\braket{\tilde{\omega}_{p}|\tilde{\omega}_{p}}}
 \mathrm{Tr}\left[\hat{J}^{-1/2}\hat{\varepsilon}\hat{J}^{-1/2}\hat{H}\right]\notag\\
 & = -\frac{1}{2}\sin(2\xi)\left(\frac{1}{\beta} - \frac{1}{\alpha}\right)\left(\frac{1}{\phi} - \frac{1}{\theta}\right). \label{eq:k-diff}
\end{align}
Equations (\ref{eq:k-rat}) and (\ref{eq:k-diff}) are enough to determine 
\begin{align}
	K_{p} = -\frac{1}{2}\sin(2\xi)\left(\frac{1}{\beta} - \frac{1}{\alpha}\right)\left(\frac{1}{\phi} - \frac{1}{\theta}\right) \frac{\omega_{p}^2}{\omega_{p}^2 - \omega_{r}^2},\label{eq:Kp}\\
	K_{r} = -\frac{1}{2}\sin(2\xi)\left(\frac{1}{\beta} - \frac{1}{\alpha}\right)\left(\frac{1}{\phi} - \frac{1}{\theta}\right) \frac{\omega_{r}^2}{\omega_{p}^2 - \omega_{r}^2}\label{eq:Kr}.
\end{align}
Note that  Eqs.~(\ref{eq:Kp}) and (\ref{eq:Kr}) are consistent with the
three requirements of rattlebacks: $\xi \neq 0$, $\alpha \neq \beta$,
and $\theta \neq \phi$. 
Equations (\ref{eq:Kp}) and (\ref{eq:Kr}) are shown to be equivalent to the
corresponding expressions Eq.~(42a,b) in Garcia and Hubbard
\cite{GarciaHubbard1988} although their expressions look quite involved.
These results also show that
\begin{equation}
K_{p}K_{r} > 0 \quad \text{and hence} \quad \overline{T}_{p}\overline{T}_{r} <0,
\end{equation}
namely, the torques generated by the pitching and the rolling always
have opposite signs to each other.
\subsubsection{Typical rattleback parameters}
Typical rattleback parameters fall in the region that satisfies
the following two conditions: 
(1) the skew angle is small,
\begin{equation}
 |\xi| \ll 1,
\end{equation}
and (2) the pitch frequency is higher than the roll frequency.  Under
these conditions, the modes $p$ and $r$ of Eq.~(\ref{eq:def-omgpr})
correspond to the pitching and the rolling oscillations respectively, and 
\begin{equation}
\omega_{p}^2 \approx  \omega_{p0}^2,\qquad
\omega_{r}^2 \approx  \omega_{r0}^2
\label{eq:omg_pr-approx-omgpr0} 
\end{equation}
in accord with the inequality (\ref{ineq:omega_pr}) \cite{Note1}.
From Eqs.~(\ref{eq:def-kpr}), (\ref{eq:Kp}), and (\ref{eq:Kr}), the signs
of the asymmetric torque coefficients and the averaged torques for
typical rattlebacks are given by
\begin{equation}
K_{p}>0 \quad \text{and}\quad K_{r}>0, \label{eq:sign-typ-k}
\end{equation}
and 
\begin{equation}
\overline{T}_{p}<0 \quad \text{and}\quad \overline{T}_{r}>0,
\end{equation}
by noting $\xi<0$, $\alpha > \beta$, $\theta >\phi$.

The fact that $\omega_{p0}>\omega_{r0}$ for a typical rattleback
means that the shape factor, $1/\phi-1$ or $1/\theta-1$, 
contributes much more  than the inertial factor, $1/\alpha$
or $1/\beta$, in 
Eqs.~(\ref{def:omega_p0}) and (\ref{def:omega_r0}) although these two factors compete, i.e.
$1/\phi-1>1/\theta-1$ and $1/\alpha<1/\beta$. This is a
typical situation because the two curvatures of usual rattlebacks are
markedly different, i.e., $\phi \ll \theta < 1$ as can be seen in
Fig.~\ref{fig:notation}(c).
Moreover, we can show that the pitch frequency is always higher for an
ellipsoid with a uniform mass density whose surface is given by $x^2/c^2
+ y^2/b^2+ z^2/a^2 = 1\ (b^2 > c^2 > a^2)$. This also holds for a
semi-ellipsoid for $b^2 > c^2 > (5/8)a^2$, where the co-ordinate system
is the same as the ellipsoid.
%

%
\subsubsection{Time for reversal}

Now we study the time evolution of the \textit{spin} $n$ defined as the
vertical component of the angular velocity
\begin{equation}
   n \equiv \bm{u}\cdot\bm{\omega},
\end{equation}
assuming that the expressions for the asymmetric torque coefficients,
$K_p$ and $K_r$, obtained above are valid even when $\omega_{z}\ne 0$.
We consider the quantities $\overline{n}$, $\overline{E}_p$, and
$\overline{E}_r$, averaged over the time scale much longer than the
oscillation periods, yet much shorter than the time scale for spin 
change.
Then, these averaged quantities should follow the following evolution equations:
\begin{align}
 I_{\textrm{eff}} \frac{d\overline{n}(t)}{dt} &= - K_{p} \overline{E}_{p}(t) + K_{r} \overline{E}_{r}(t) , \label{eq:diff-ne-1}\\
 \frac{d \overline{E}_{p}(t)}{dt} & =  K_{p} \overline{n}(t) \overline{E}_{p}(t),\label{eq:diff-ne-2}\\
 \frac{d \overline{E}_{r}(t)}{dt} & = - K_{r} \overline{n}(t) \overline{E}_{r}(t) \label{eq:diff-ne-3}.
\end{align}
Here, $I_{\textrm{eff}}$ is the effective moment of inertia  around $\bm{u}$ under
the existence of the oscillations, and is assumed to be constant; it should be close to $I_{zz}$.
As can be seen easily, the total energy $E_{\textrm{tot}}$ defined by
\begin{equation}
 E_{\textrm{tot}} \equiv \frac{1}{2} I_{\textrm{eff}} \overline{n}(t)^2 + \overline{E}_{p}(t) + \overline E_{r}(t)
\end{equation}
is conserved. It can be seen that there is another invariant, 
\begin{equation}
C_{I} \equiv \frac{1}{K_{p}}\ln \overline{E}_{p} + \frac{1}{K_{r}}\ln \overline{E}_{r},\label{eq:casimir}
\end{equation}
which has been discussed in connection with a Casimir invariant
\cite{MoffattTokieda2008, Yoshida2016}. With these two
conservatives, general solutions of the three-variable system
(\ref{eq:diff-ne-1})--(\ref{eq:diff-ne-3}) should be periodic.

Let us consider the case where the spin is positive at $t=0$ and the sum of the oscillation energies are small compared to the spinning energy:
\begin{equation}
 \overline{n}(0) \equiv n_i >0, \quad 
\overline{E}_{p}(0) + \overline{E}_{r}(0) \ll \frac{1}{2} I_{\textrm{eff}} n_{i}^2.
\end{equation}
For a typical rattleback, the pitching develops and the rolling decays as long
as $\overline n>0$ as can be seen from Eqs.~(\ref{eq:sign-typ-k}), (\ref{eq:diff-ne-2}) and (\ref{eq:diff-ne-3}). Thus the rolling is irrelevant and can be ignored, i.e., $\overline E_r(t) = 0$, to estimate the time for reversal. Then we can derive the equation
\begin{equation}
 \frac{d\overline{n}(t)}{dt} = -\frac{K_{p}}{2}\left(n_0^2- \overline{n}(t)^2\right) \label{eq:diff-n},
\end{equation}
where the constant  $n_0>0$ is defined by
\begin{equation}
  \frac{1}{2}I_{\textrm{eff}} n_{0}^2 \equiv  E_{\textrm{tot}}.
\end{equation}
This can be easily solved as
\begin{equation}
\overline{n}(t) = n_{0}\frac{(n_{0} + n_{i})\exp(-n_{0}K_{p}t) - (n_{0} -n_{i}) }{(n_{0} + n_{i})\exp(-n_{0}K_{p}t) + (n_{0} -n_{i})}, 
\label{eq:gh-solution-p}
\end{equation}
and we obtain the time for reversal $t_{rGH+}$ for the $n_{i} > 0$ case as
\begin{equation}
 t_{rGH+} = \frac{1}{n_0 K_p} \ln\left(\frac{n_{0}+n_{i}}{n_0 - n_i}\right),
\label{eq:trgh-p}
\end{equation}
by just setting $\overline{n}=0$ in Eq.~(\ref{eq:gh-solution-p}).

Similarly, in the case of $n_i<0$, only the rolling develops and the pitching is irrelevant, thus we obtain $\overline{n}(t)$ and the time for reversal $t_{rGH-}$ as
\begin{equation}
\overline{n}(t) = -n_{0}\frac{(n_{0} + |n_{i}|)\exp(-n_{0}K_{r}t) - (n_{0} - |n_{i}|) }{(n_{0} + |n_{i}|)\exp(-n_{0}K_{r}t) + (n_{0} - |n_{i}|)} 
\label{eq:gh-solution-m}
\end{equation}
and 
\begin{equation}
 t_{rGH-} =\frac{1}{n_0 K_r} \ln\left(\frac{n_0+|n_i|}{n_0 - |n_i|} \right). \label{eq:trgh-m}
\end{equation}
Equations (\ref{eq:trgh-p}) and (\ref{eq:trgh-m}) are Garcia-Hubbard formulas for the times for reversal \cite{GarciaHubbard1988}. 

From the expressions of $K_{p}$ and $K_{r}$ given by Eqs.~(\ref{eq:Kp})
and (\ref{eq:Kr}), we immediately notice that (1) the time for reversal
is inversely proportional to the skew angle $\xi$ in the small skewness
regime, and (2) the ratio of the time for reversal $t_{rGH-}/t_{rGH+}$
is simply given by the squared ratio of the pitch frequency to the roll
frequency $\omega_{p}^2 / \omega_{r}^2$, provided initial values $n_{0}$
and $n_{i}$ are the same except their signs.

For a typical rattleback, $\omega_{p}^2 \gg \omega_{r}^2$, thus $t_{rGH+} \ll t_{rGH-}$, i.e., the time for reversal is much shorter in the case of $n_{i}>0$ than in the case of $n_{i}<0$. Thus we call the spin direction of $n_{i} > 0$ the \textit{unsteady direction} \cite{GarciaHubbard1988}, and that of $n_i <0$ the \textit{steady direction}.

In the small skewness regime, this ratio of the squared frequencies is estimated as
\begin{align}
\frac{\omega_{p}^2}{\omega_{r}^2} \approx \frac{\omega_{p0}^2}{\omega_{r0}^2}
 = \frac{\beta}{\alpha}\frac{1/\phi - 1}{1/\theta -1}. \label{eq:ratomg}
\end{align}
This becomes especially large as $\theta$ approaches $1$ or as $\phi$ approaches 0, namely, as the smaller radius of principal curvature approaches $a$, or as the larger radius of principal curvature becomes much larger than $a$. 
We remark that both of the inertial parameters $\alpha$ and $\beta$ are larger than $1$ by definition Eq.~(\ref{eq:def-abg}), and cannot be arbitrarily large for a typical rattleback.

Let us consider these two limiting cases: $\phi\to 0$ and $\theta \to 1$ with $|\xi| \ll 1$. In the case of $\phi \to 0$, 
\begin{equation}
 K_{p} \to \infty,\quad
 K_{r} \to (-\xi) \left(\frac{1}{\beta} - \frac{1}{\alpha}\right)\frac{\alpha}{\beta}\left(\frac{1}{\theta} - 1\right),
\end{equation}
thus the time for reversal $t_{rGH-}$ remains constant while $t_{rGH+}$ approaches $0$. In the case of $\theta \to 1$,
\begin{equation}
 K_{p} \to (-\xi) \left(\frac{1}{\beta} - \frac{1}{\alpha}\right)\left(\frac{1}{\phi} - 1\right),\quad
 K_{r} \to 0,
\end{equation}
and thus $t_{rGH+}$ remains constant while $t_{rGH-}$ diverges to infinity,
i.e., the negative spin rotation never reverses.

\section{Simulation}
\label{sec:simulation}

We perform numerical simulations for the times for the first spin reversal and compare them with Garcia-Hubbard formulas (\ref{eq:trgh-p}) and (\ref{eq:trgh-m}).

\subsection{Shell-dumbbell model}
To consider a rattleback whose inertial and geometrical parameters can be set separately, we construct a simple model of the rattleback, or the \textit{shell-dumbbell model}, which consists of a light shell and two dumbbells: the light shell defines the shape of the lower part of the rattleback and the dumbbells represent the masses and the moments of inertia. The shell is a paraboloid given by Eq.~(\ref{eq:def-z}). The dumbbells consist of couples of weights, $m_{x}/2$ and $m_{y}/2$, fixed at $(\pm r_{x},0,0)$ and $(0,\pm r_{y},0)$ in the body-fixed co-ordinate, respectively [Fig.~\ref{fig:notation}(c)]. Then the total mass is 
\begin{equation}
 M = m_{x} + m_{y}
\end{equation}
and the inertia tensor is diagonal with its principal moments 
\begin{align}
	I_{xx} &= m_{y}r_{y}^{2},\quad I_{yy} = m_{x}r_{x}^{2}, \\
	I_{zz} &= m_{y}r_{y}^{2} + m_{x}r_{x}^{2}.
\end{align}
Note that the simple relation 
\begin{equation}
	I_{zz} = I_{xx} + I_{yy}
\end{equation}
holds for the shell-dumbbell model. We define 
\begin{equation}
 f_{sd} \equiv I_{yy}/I_{zz},
\end{equation}
then the dimensionless parameters $\alpha,\,\beta,$ and $\gamma$ defined by Eq.~(\ref{eq:def-abg}) are given by,
\begin{equation}
	\gamma = I_{zz}/Ma^{2}, \, \alpha = (1-f_{sd})\gamma + 1, \, \beta = f_{sd}\gamma + 1.
\end{equation}
The parameter $f_{sd}$ satisfies $0<f_{sd}<0.5$, since we have assumed $\alpha > \beta$.

The shell-dumbbell model makes it easier to visualize an actual object represented by the model with a set of parameters, and is used in the following simulations for determining the parameter ranges.
\subsection{Methods}
\label{subsec:method}
%
\begin{table*}
	\caption{\label{tab:parameters} Two sets of parameters used in the simulations: GH used by Garcia and Hubbard \cite{GarciaHubbard1988} and SD for the present shell-dumbbell model. For SD, the parameter values are chosen randomly from the ranges shown, and averages and/or distributions of simulation results are presented.}
\begin{ruledtabular}
	\begin{tabular}{lccccccc}
	 &$\gamma$ & $f_{sd}$ & $\alpha,\ \beta$ & $\theta$ & $\phi$ & $-\xi$ (deg) \\ \midrule
	 GH & $12.28$ &---& 13.04, 1.522 & 0.6429& 0.0360 & 1.72\\
	 SD & $[5,15]$ &$[0.05,0.15]$ &---& $[0.6,0.95]$ & [0.01,0.1]& (0,6] \\
	\end{tabular}
\end{ruledtabular}
\end{table*}
%
\begin{figure}
\centering \includegraphics[width=7.5cm]{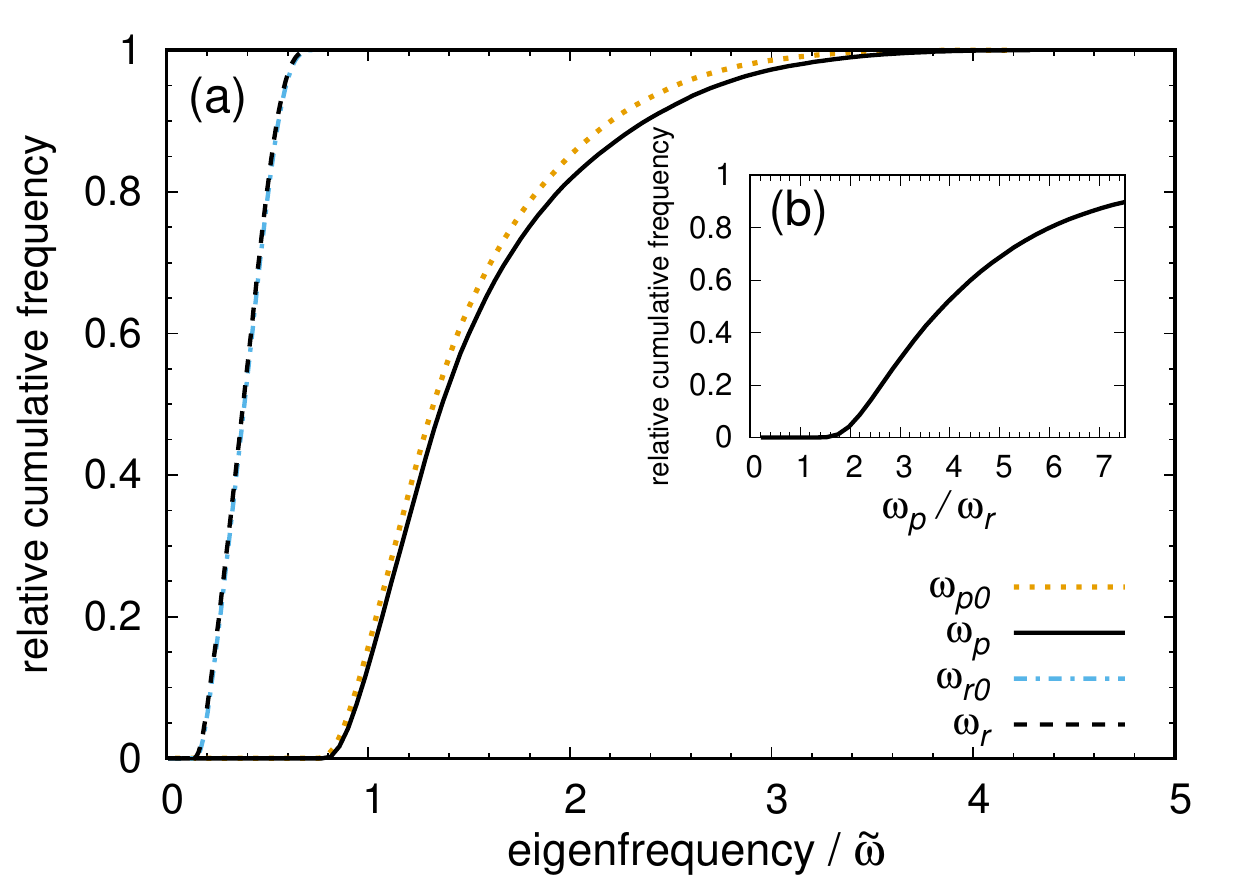}
\caption{\label{fig:omega-dist}(color online) (a) Cumulative
distributions of the pitch and the roll frequencies for the parameter
set SD in Table~\ref{tab:parameters}; $\omega_{p} \text{~and~}
\omega_{r}$ of Eq.~(\ref{eq:def-omgpr}) and their zeroth order
approximation $\omega_{p0}$ and $\omega_{r0}$ by 
Eqs.~(\ref{def:omega_p0}) and (\ref{def:omega_r0})} are shown. The
inset shows the cumulative distribution of $\omega_{p}/\omega_{r}$. The
number of samples is $10^{6}$.
\end{figure}
\begin{figure}
\centering
  \includegraphics[width=7.5cm]{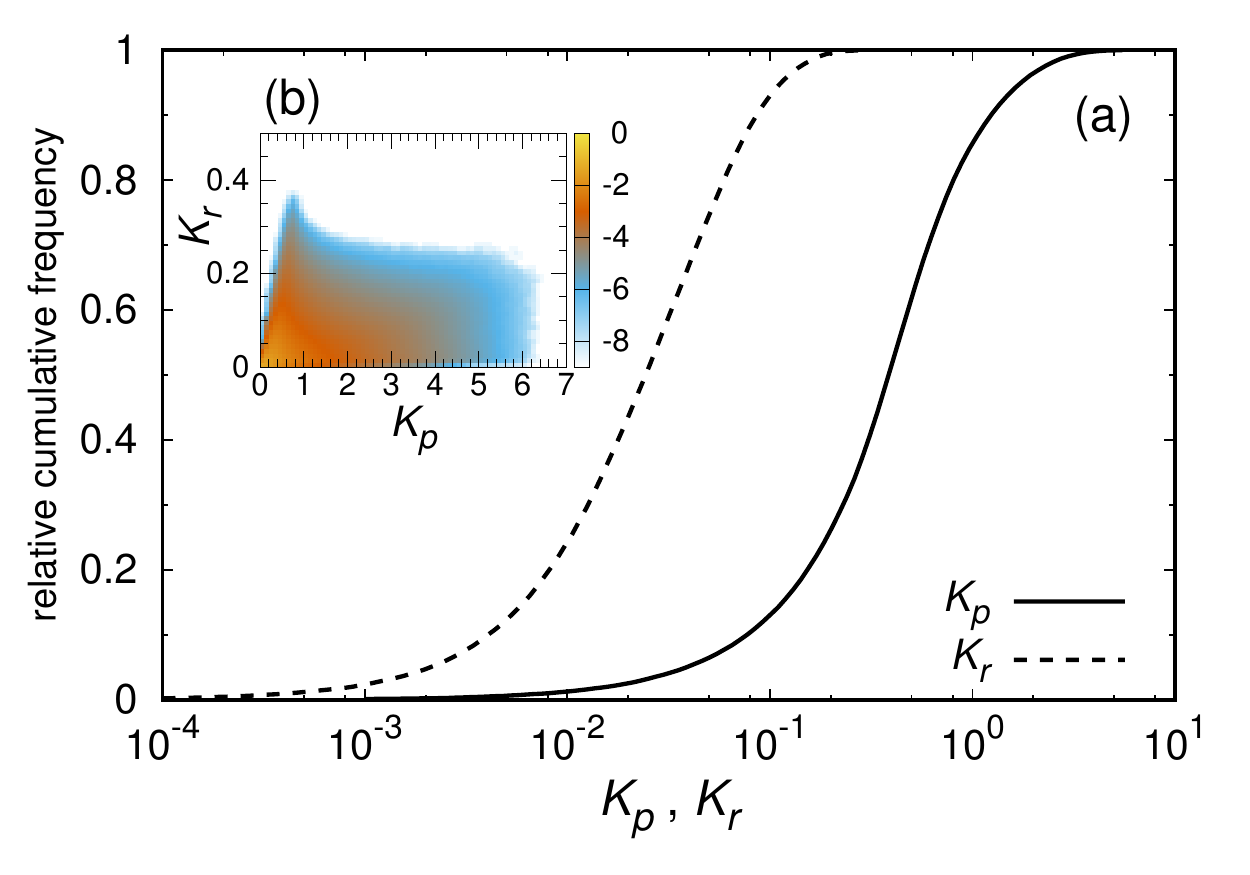}
\caption{\label{fig:k-dist}(color online) (a) Cumulative distributions of the  asymmetric torque coefficients $K_{p}$ and $K_{r}$ for SD (Table \ref{tab:parameters}). The number of samples is ${10}^{5}$. (b) A 2D color plot for the distribution of ($K_{p}$,\,$K_{r}$). The color code shown is in the logarithmic scale for the relative frequency $P(K_{p},\,K_{r})$, i.e., $-9 \le \log_{10}P(K_{p},\,K_{r}) \le 0$. The number of samples is $10^{8}$.}
\end{figure}
%
The equations of motion (\ref{eq:diff-u}) and (\ref{eq:diff-omega}) with the contact point  conditions (\ref{eq:def-z}) and (\ref{eq:def-xy}) are numerically integrated by the fourth-order Runge-Kutta method with an initial condition $\bm{\omega}(0)$ and $\bm{u}(0)$.
In the simulations, we take
\begin{equation}
 \bm{u}(0) = (0,0,-1)^{t} \label{eq:sim-ic-1}
\end{equation}
and specify $\bm{\omega}(0)$ as
\begin{equation}
 \bm{\omega}(0) = (|\omega_{xy0}|\cos\psi,\  |\omega_{xy0}|\sin\psi,\  -n_{i}) \label{eq:sim-ic-2}
\end{equation}
in terms of $|\omega_{xy0}|$, $\psi$, and $n_{i}$. According to the simplified dynamics (\ref{eq:diff-ne-1})--(\ref{eq:diff-ne-3}), the irrelevant mode of oscillation does not affect the dynamics sensitively as long as the relevant mode exists and the initial spin energy is much larger than the initial oscillation energy. Thus we choose $\ket{\omega(0)}= (\omega_{x0}, \omega_{y0})^{t}$ in the direction of the relevant eigenmode,
\begin{equation}
\psi = \psi_{p} \text{ for } n_{i}>0,\quad\text{and}\quad\psi = \psi_{r} \text{ for } n_{i}<0, \label{eq:sim-ic-3}
\end{equation}
where $\psi_{p}$ and $\psi_{r}$ are the angles of the eigenvectors $\ket{\omega_{p}}$ and $\ket{\omega_{r}}$ from the $x$-axis, respectively.

Numerical results are presented in the unit system where $M$, $a$, and 
\begin{equation}
\tilde{t} \equiv 1/\tilde{\omega} \equiv \sqrt{a/g}
\end{equation}
as units of mass, length, and time. The size of the time step for the numerical integration is taken to be $0.002\,\tilde{t}$.
In numerics, we determine the time for reversal $t_{r}$ by the time at which $n=\bm{\omega}\cdot\bm{u}$ becomes zero for the first time, and they are compared with Garcia-Hubbard formulas  (\ref{eq:trgh-p}) and (\ref{eq:trgh-m}); $n_0$ is determined as 
\begin{equation}
	\frac{\gamma n_{0}^{2}}{2} = \frac{1}{2}(
	\alpha\omega_{x0}^2 + \beta\omega_{y0}^2 + \gamma\omega_{z0}^2),
	\label{eq:def-n0}
\end{equation}
assuming $I_{\mathrm{eff}} = I_{zz}$ at $t=0$.
Here the potential energy $U(\bm{u})$ is set to zero where $\bm{u}(0) = (0,0,-1)^{t}$.

The parameters used in the simulations are listed in Table~\ref{tab:parameters}.  
For the parameter set SD, the ranges are shown. When numerical results are plotted against $K_{p}$ or $K_{r}$, given by Eqs.~(\ref{eq:Kp}) or (\ref{eq:Kr}), respectively, sets of parameters are chosen randomly from the ranges until resulting $K_{p}$ or $K_{r}$ falls within the range of $\pm0.1\%$ of a target value. 
The ranges of SD are chosen to meet the following two conditions: (1)
$0<\phi \ll \theta <1$, $\beta < \alpha$, and $|\xi| \ll 1$ and (2) the
pitch frequency should be higher than the roll frequency. As argued
in Sec.~\ref{subsec:GHtheory}, usual rattlebacks such as one in
Fig.~\ref{fig:notation}(a) satisfy these two
conditions. Figure \ref{fig:omega-dist} shows the cumulative distributions
for the eigenfrequencies $\omega_{p}$ and $\omega_{r}$, and their
approximate expressions $\omega_{p0}$ and $\omega_{r0}$ for the
parameter set SD; it shows $(\omega_{p}/\omega_{r}) > 1.3$ in accordance
with the condition (2).

The parameter set GH gives $K_{p} = 0.553$ and $K_{r}=0.0967$, and the
distributions of $K_{p}$ and $K_{r}$ for SD are shown in
Fig.~\ref{fig:k-dist}, where one can see $K_{p}\gg K_{r}$. From
Eq.~(\ref{eq:k-rat}), this corresponds to $\omega_{p}^2 \gg
\omega_{r}^2$, i.e., the pitch frequency is significantly faster than
the roll frequency. Consequently, the time for reversal is much shorter
for the unsteady direction $n_{i}>0$, where the pitching is induced,
than for the steady direction $n_{i}<0$, where the rolling is
induced. We denote the time for reversal for the unsteady direction as
$t_{ru}$ and that for the steady direction as $t_{rs}$ when we consider
a specific spinning direction.
\subsection{Results}
\subsubsection{General behavior for the parameter set GH}
%
\begin{figure}
 \includegraphics[width=8cm]{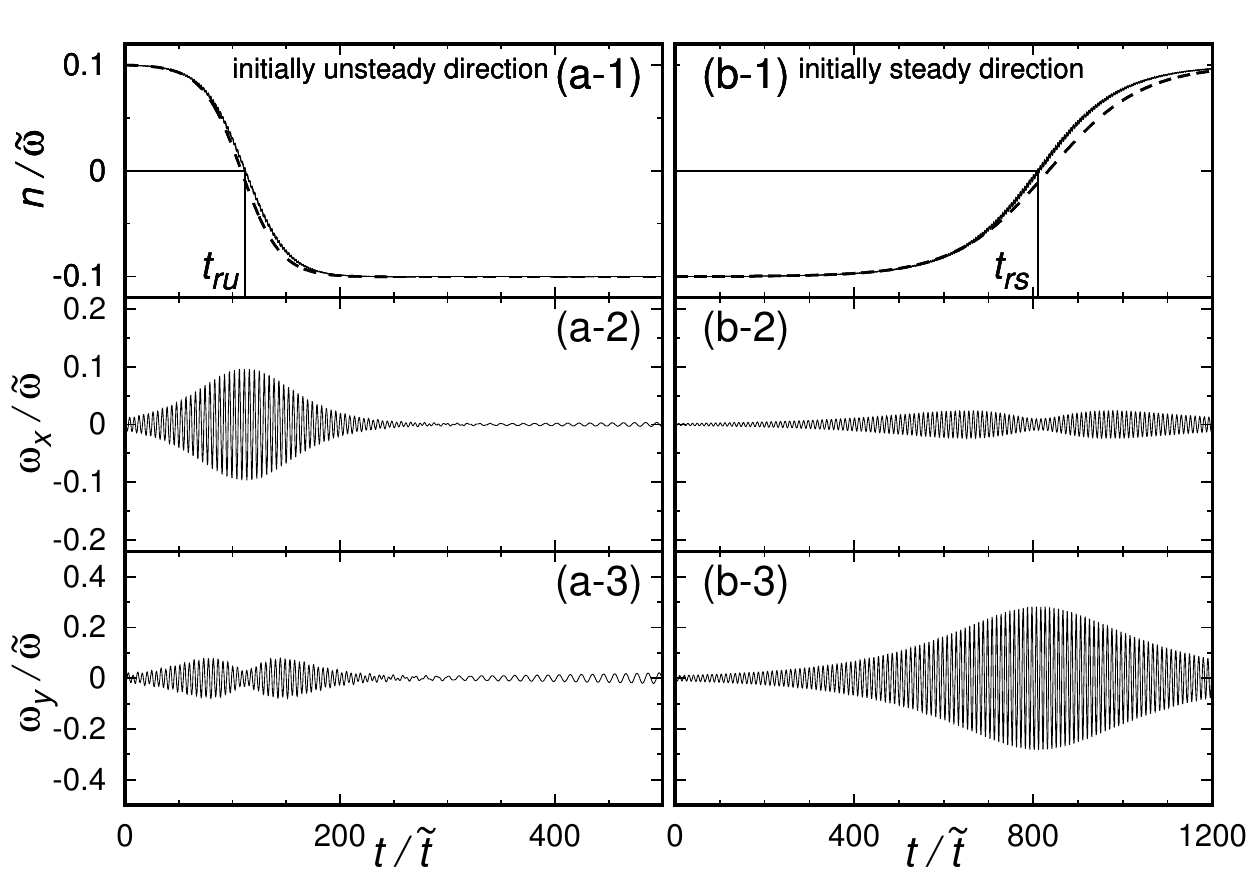}
\caption{\label{fig:gh-t-n} A typical spin evolution and the corresponding $\omega_{x}$ and $\omega_{y}$ for GH (Table~\ref{tab:parameters}). (a) The case of the initial spin in the unsteady direction. The initial condition is specified by Eqs.~(\ref{eq:sim-ic-1})--(\ref{eq:sim-ic-3}) with $n_{i} = 0.1\,\tilde{\omega}$ and $|\omega_{xy0}|=0.01\,\tilde{\omega}$. (b) The case of the initial spin in the steady direction with $n_{i} = -0.1\,\tilde{\omega}$ and $|\omega_{xy0}|=0.01\,\tilde{\omega}$. The dashed lines in (a-1) and (b-1) show Garcia and Hubbard's solution $\overline{n}(t)$ given by Eqs.~(\ref{eq:gh-solution-p}) and (\ref{eq:gh-solution-m}), respectively.}
\end{figure}
%
In Fig.~\ref{fig:gh-t-n} we show a typical simulation result of the time evolution of the spin $n(t)$ along with the angular velocities $\omega_{x}(t)$ and $\omega_{y}(t)$ for the parameter set GH (Table~\ref{tab:parameters}) in the case of the unsteady direction $n_i > 0$ (a), and the steady direction $n_i < 0$ (b).

Figure \ref{fig:gh-t-n}(a-1) shows that the spin $n$ changes its sign from positive to negative  at $t_{ru}\approx 112\,\tilde{t}$, and Fig.~\ref{fig:gh-t-n}(b-1) shows the spin $n$ changes its sign from negative to positive at $t_{rs} \approx 810\,\tilde{t}$. Garcia and Hubbard's solutions $\overline{n}(t)$ of Eqs.~(\ref{eq:gh-solution-p}) and (\ref{eq:gh-solution-m}) are shown by the dashed lines in Figs.~\ref{fig:gh-t-n}(a-1) and (b-1), respectively; they are in good agreement with the numerical simulations. 

The angular velocities $\omega_{x}$ and $\omega_{y}$ oscillate in much
shorter time scale, and their amplitudes evolve differently depending on
the spin direction. In the case of Fig.~\ref{fig:gh-t-n}(a), where the
positive initial spin reverses to negative, the amplitude of
$\omega_{x}$ becomes large and reaches its maximum around $t_{ru}$; the
amplitude of $\omega_{y}$ also becomes large around both sides of
$t_{ru}$ but shows the local minimum at $t_{ru}$. Both $\omega_{x}$
and $\omega_{y}$ oscillate at the pitch frequency $\omega_{p} \approx
1.44\,\tilde{\omega}$. In the case of Fig.~\ref{fig:gh-t-n}(b) where the
negative spin reverses to positive, the situation is similar but the
amplitude of $\omega_{y}$ reaches its maximum around $t_{rs}$, and
$\omega_{x}$ and $\omega_{y}$ oscillate at the roll frequency
$\omega_{r} \approx 0.602\,\tilde{\omega}$.

These features can be understood based on the analysis in the previous
section as follows. The positive spin induces the pitching, which is
mainly represented by $\omega_{x}$ because the eigenvector of the pitching
$\ket{\omega_{p}}$ is nearly parallel to the $x$ axis, i.e., $\psi_{p}
\approx -17^{\circ}$. Likewise, the negative spin induces the rolling,
mainly represented by $\omega_{y}$, because $\psi_{r} \approx
88^{\circ}$. The local minima of the amplitude for $\omega_{y}$ in
Fig.~\ref{fig:gh-t-n}(a-3), or $\omega_{x}$ in
Fig.~\ref{fig:gh-t-n}(b-2), at the times for reversal are tricky; it
might mean that the eigenvector of the pitching (rolling) deviates 
more from the $x$ axis ($y$ axis) for $\omega_{z} \neq 0$ than
that for $\omega_{z} = 0$; as a result, the pitching (rolling) mode has a 
larger projection on the $y$ axis ($x$ axis) for $\omega_{z} \neq 0$.

 Note that for given $|n_{i}|$, the maximum value of $\omega_{y}$ in
 Fig.~\ref{fig:gh-t-n}(b-3) is larger than that of $\omega_{x}$ in
 (a-2). This is due to $\alpha \gg \beta$; the oscillation energy
 around zero spin for the both cases should be the same, which gives $\alpha \overline{{\omega}_{x}^2}\approx \beta \overline{{\omega}_{y}^2}$ thus
 $\sqrt{\overline{\omega_{x}^2}} < \sqrt{\overline{\omega_{y}^2}}$.
\subsubsection{Simulations with the parameter set SD}
\begin{figure}
\includegraphics[width=8.4cm]{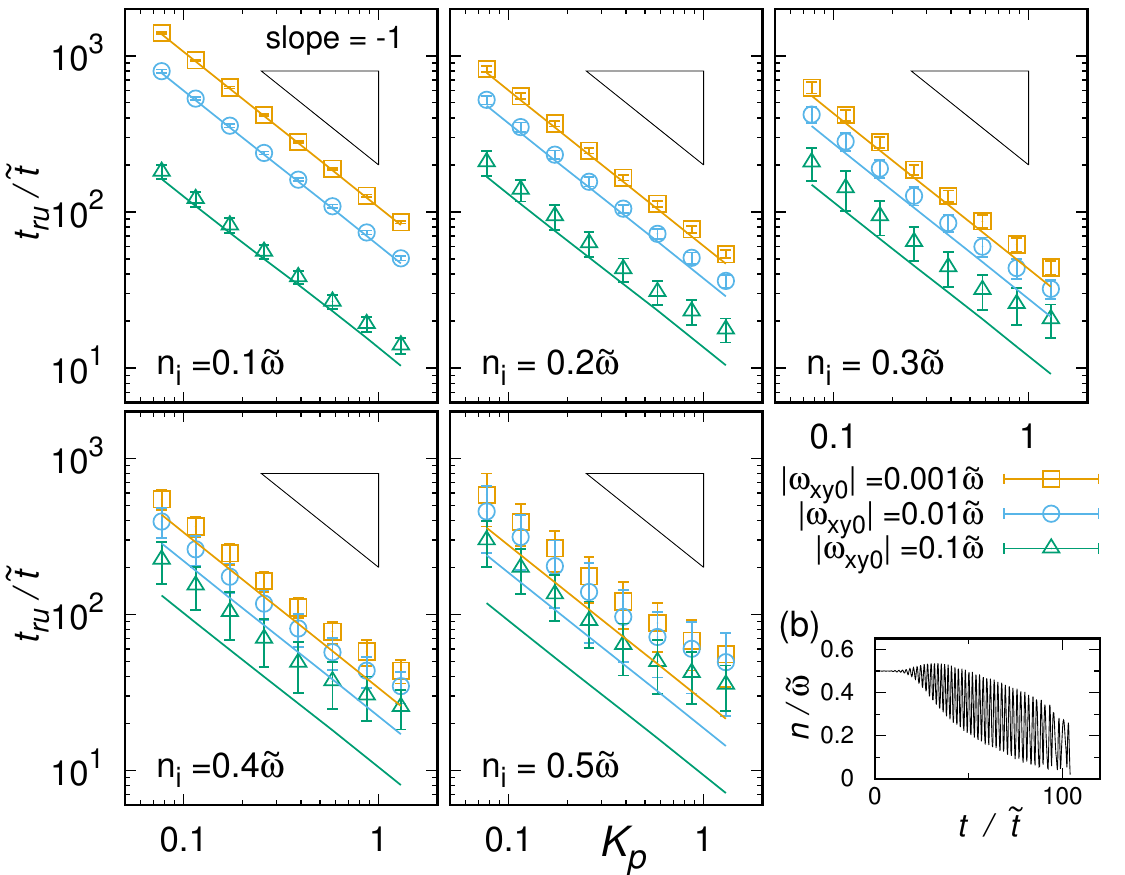} \caption{(color online)
\label{fig:kptr} (a) Time for reversal of the unsteady direction
$t_{ru}$ for the parameter set SD (Table~\ref{tab:parameters}) as a
function of the asymmetric torque coefficient $K_{p}$ in the logarithmic
scale. The error bars indicate one standard deviation of $1000$ samples
for each data point. The solid lines are $t_{rGH+}$ given by
Eq.~(\ref{eq:trgh-p}), calculated using the mean values of $n_{0}$. (b)
A typical spin evolution with $n_{i} = 0.5\,\tilde{\omega},\
|\omega_{xy0}|=0.01\,\tilde{\omega}$. The parameter set GH is used.}
\end{figure}

We present detailed results of the simulations for the ranges of the parameters given by SD in Table \ref{tab:parameters}.
\paragraph{Unsteady initial spin direction $(n_{i}>0)$.}
In this case, the system behaves basically as we expect from the Garcia-Hubbard formula unless the initial spin or oscillation is too large.
Figure \ref{fig:kptr} shows the time for reversal $t_{ru}$ as a function of $K_{p}$ when spun in the unsteady direction. The results are plotted against $K_{p}$ by the procedure described in Sec.~\ref{subsec:method}. 

When the initial spin $n_{i}$ is  $n_{i} \lesssim 0.2\,\tilde{\omega}$ with $|\omega_{xy0}|=0.001\tilde{\omega},\, 0.01\tilde{\omega}$, $t_{ru}$ is in good agreement with the Garcia-Hubbard formula $t_{rGH+}$ of Eq.~(\ref{eq:trgh-p}), i.e., almost inversely proportional to $K_{p}$ with small scatter around the average. For a given $n_i$, as the initial oscillation amplitude $|\omega_{xy0}|$ becomes large, the standard deviations of $t_{ru}$ become large, and the average of $t_{ru}$ deviates  upward from the Garcia-Hubbard formula $t_{rGH+}$, which is derived with the small amplitude approximation of $\omega_{x}$ and $\omega_{y}$. For larger $n_i$, $t_{rGH+}$ also underestimates $t_{ru}$, as already noted by Garcia and Hubbard \cite{GarciaHubbard1988} for the parameter set GH. The underestimation can be also seen in Fig.~\ref{fig:gh-t-n}(a-1), where one can see that Garcia and Hubbard's solution $\overline{n}(t)$ of Eq.~(\ref{eq:gh-solution-p}) changes its sign earlier than the simulation. 

For $n_i \gtrsim 0.4\,\tilde{\omega}$, $t_{ru}$ deviates notably upward from the Garcia-Hubbard formula $t_{rGH+}$. As $n_i$ increases, the average of $t_{ru}$ increases and the standard deviations become large. Figure \ref{fig:kptr}(b) shows a typical spin evolution with $n_{i} = 0.5\,\tilde{\omega}$. The spin oscillates widely at the pitch frequency, which is qualitatively different from typical spin behaviors at small $n_{i}$ and from Garcia and Hubbard's solution $\overline{n}(t)$ of Eq.~(\ref{eq:gh-solution-p}) as in Fig.~\ref{fig:gh-t-n}(a-1). In this region, the Garcia-Hubbard formula is no longer valid.
%
\paragraph{Steady initial spin direction $(n_{i}<0)$.}
%
\begin{figure*}
\includegraphics[width=16.8cm]{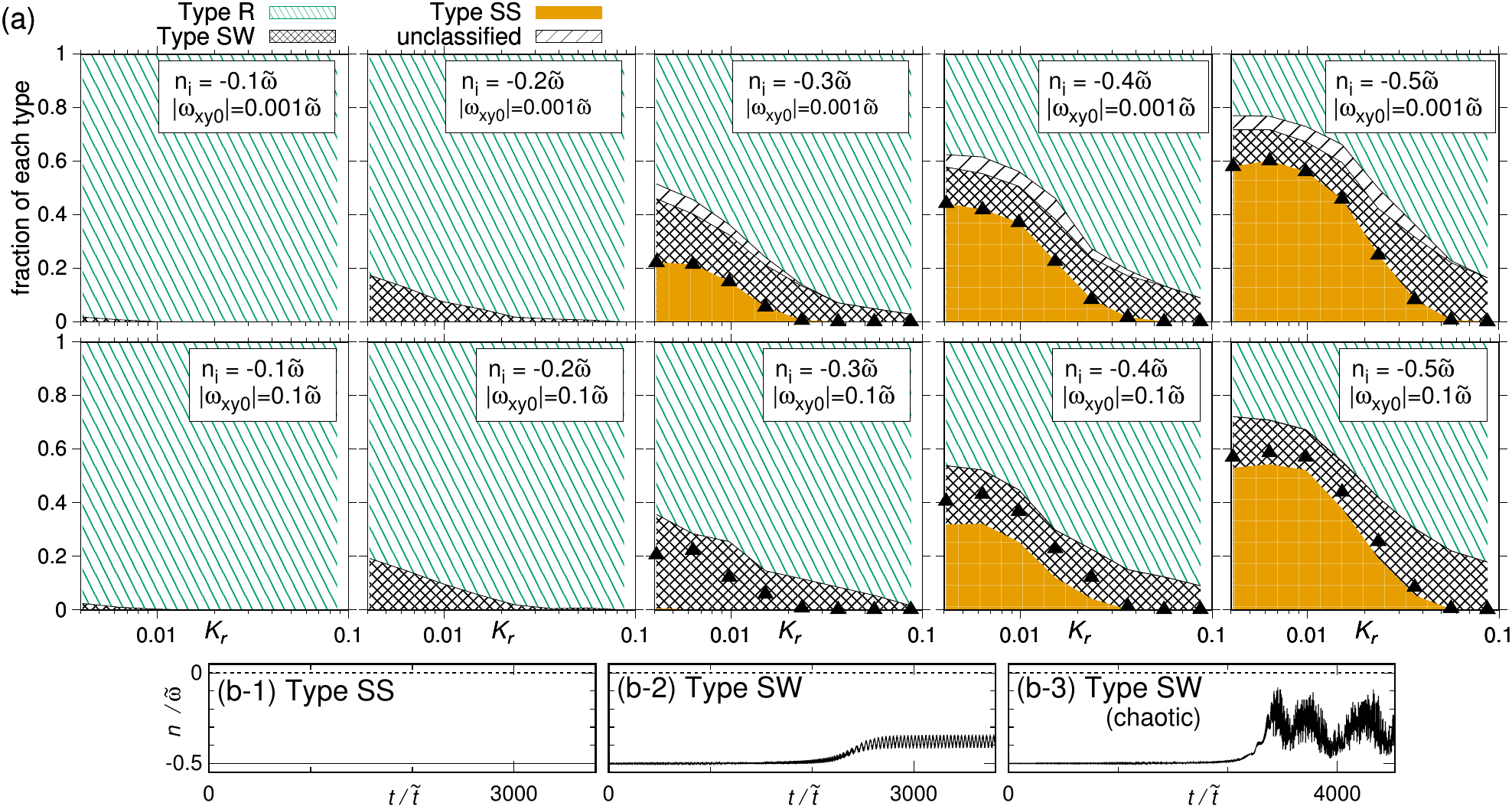} \caption{(color online)
\label{fig:krtr}(a) Fractions of Types R, SS, and SW for the steady
direction for eight values of $K_{r}$ with various initial conditions
$|\omega_{xy0}| \text{ and } n_{i}$. Parameters are randomly chosen from
SD (Table~\ref{tab:parameters}). The number of the samples is 1000 for
each $K_{r}$. Filled triangles show the fractions of samples whose
$|n_{c1}|$ is smaller than $|n_{i}|$.  (b) Typical spin evolutions of a
Type SS sample (b-1) and a Type SW sample (b-2), along
with an example of ``chaotic" oscillation (b-3) found for $K_{r} = 0.0041$
with $n_{i}=-0.5\,\tilde{\omega}$.}
\end{figure*}
\begin{figure}
\includegraphics[width=8.4cm]{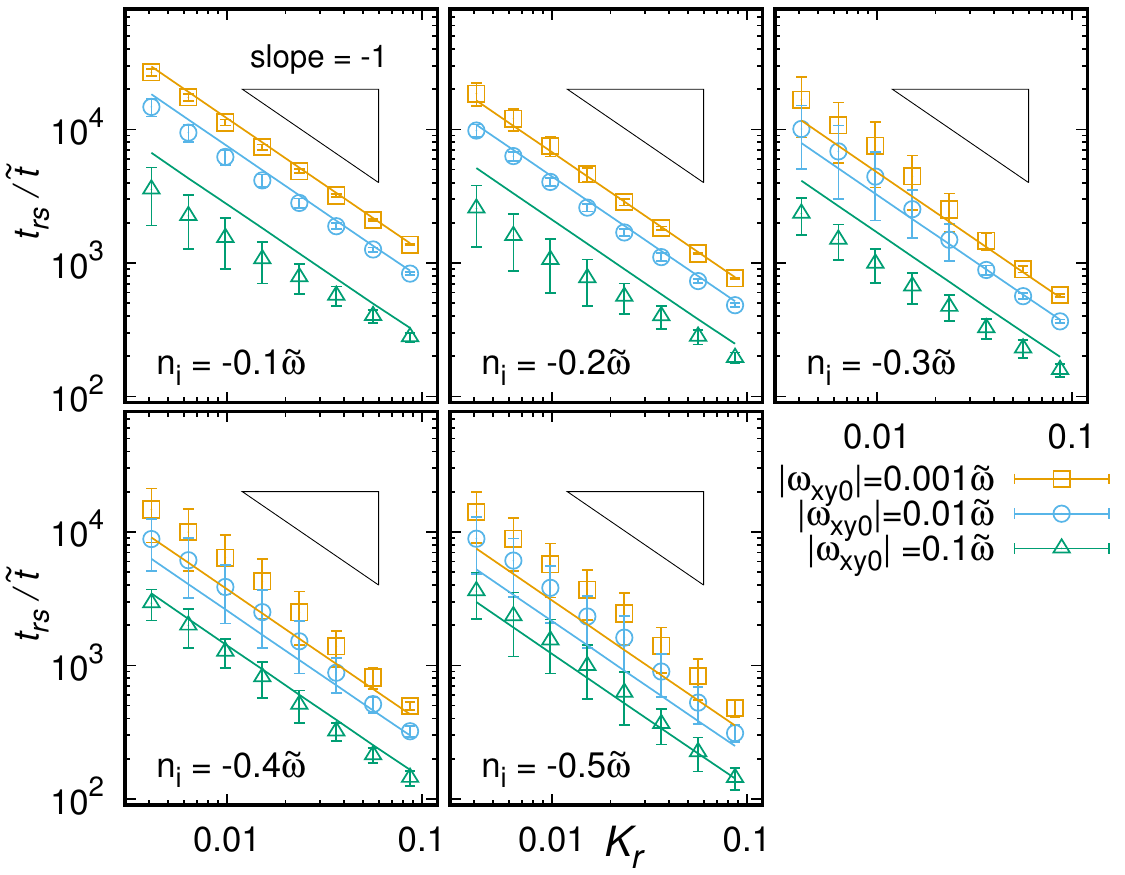} \caption{(color online)
\label{fig:krtr-2} Time for reversal $t_{rs}$ for the steady direction
as a function of $K_{r}$ in the logarithmic scale. Each data point
represents the average with the standard deviation of Type R samples out
of $1000$ simulations from the parameter set SD
(Table~\ref{tab:parameters}).}
\end{figure}
%
Much more complicated phenomena are observed when spun in the steady
direction. When the initial spin $|n_i|$ is small enough, the spin
simply reverses as shown in Fig.~\ref{fig:gh-t-n}(b-1).  We call this
simple reversal behavior Type R. For larger $|n_i|$, however, there
appear some cases where the spin never reverses; in such cases there are
two types of behaviors: steady spinning at $n_{ss}$ (Type SS), and spin
wobbling around $n_{w}$ ($n_{ss}<n_{w}<0$, Type SW). For Type SS
samples, $n_{ss}$ is slightly less than $n_{i}$, 
i.e., $n_{ss}<n_{i}<0$, because small initial rolling decays and
its energy is converted to the spin energy. Typical spin evolutions of a
Type SS sample and a Type SW sample are shown in
Figs.~\ref{fig:krtr}(b-1) and (b-2).

Figure \ref{fig:krtr}(a) shows the $K_{r}$ dependence of the fractions of
Types R, SS, and SW for various initial conditions given by $n_{i}$ and
$|\omega_{xy0}|$. For each sample, we wait up to $t = 5t_{rGH-}$;
the spin evolution is classified as Type R if it reverses. If it does
not, the spin evolution is classified as Type SS if the initial rolling
amplitude decays monotonously, and classified as Type SW if the spin $n$
starts wobbling by the time $5t_{rGH-}$. The other samples, in 
which the rolling grows slowly yet shows no visible spin change
by the time $5t_{rGH-}$, are labeled ``unclassified" in
Fig.~\ref{fig:krtr}. Such samples may show spin reversal or spin
wobbling if we take a much longer simulation time.  Type SS appears for
$|n_i|\gtrsim 0.3\,\tilde{\omega}$ and its fraction increases as $|n_i|$
increases. The fraction is larger for smaller $K_{r}$ and smaller
$|\omega_{xy0}|$, i.e., $|\omega_{xy0}|=0.001\,\tilde{\omega}$.  Type SW
appears for $|n_i| \gtrsim 0.1\,\tilde{\omega}$ and its fraction is also
larger for smaller $K_{r}$, but stays around $0.2$ for $|n_i| \gtrsim
0.4\,\tilde{\omega}$.

Figure \ref{fig:krtr-2} shows the $K_{r}$ dependence of $t_{rs}$ 
only for the samples of Type R, which shows a spin reversal
behavior. For small $|n_{i}| \lesssim 0.2\,\tilde{\omega}$ with
$|\omega_{xy0}|=0.01\,\tilde{\omega}, 0.001\,\tilde{\omega}$, $t_{rs}$
is in good agreement with Garcia-Hubbard formula $t_{rGH-}$ of
Eq.~(\ref{eq:trgh-m}), and the average of $t_{ru}$ is almost
inversely proportional to $K_{r}$. As in the case of the unsteady
direction, the standard deviations of $t_{rs}$ become large, and the
average $t_{rs}$ deviates downward from $t_{rGH-}$ as initial
oscillation amplitude $|\omega_{xy0}|$ becomes large. Note that
$t_{rGH-}$ tends to overestimate $t_{rs}$, in contrast to the case of
the unsteady direction, where $t_{rGH+}$ underestimates $t_{ru}$. This
has also been noted by Garcia and Hubbard \cite{GarciaHubbard1988} for
the parameter set GH, and can be seen by Garcia and Hubbard's solution
$\overline{n}(t)$ in Fig.~\ref{fig:gh-t-n}(b-1). For $|n_{i}| \gtrsim
0.3\,\tilde{\omega}$, one may notice the standard deviations are large
for $K_{r}\ll 0.1$. In these cases, we find that some samples appear to
spin stably for quite a long time, i.e., several times of
$t_{rGH-}$, and then abruptly starts to reverse its sign. During the
time period $t<t_{rs}$, the rolling grows much more slowly than it
should as predicted by the theory in Sec.~\ref{sec:theory}. Such samples
make both the average and standard deviation large as
Fig.~\ref{fig:krtr-2}.

Next we consider the Type SS samples. There always exists a steady solution, $\bm{\omega}(0)=(0,0,\mathrm{const.})^{t}$ and $\bm{u}(0) = (0,0,-1)^{t}$, and Bondi \cite{Bondi1986} has shown that for the steady direction, this solution is linearly stable for $n < n_{c1}<0$, where $n_{c1} (<0)$ is given by
\begin{equation}
n_{c1}^2 \equiv \frac{g}{a}\frac{-(1-\theta)(1-\phi)}{2-(\theta+\phi) - (\alpha + \beta - \gamma)(\theta + \phi - 2\theta\phi)}. \label{eq:def-nc1}
\end{equation}
When the denominator of Eq.~(\ref{eq:def-nc1}) is positive, such a threshold does not actually exist, and the steady solution is always unstable. Note that $n_{c1}$ does not depend on $\xi$. 

In Fig.~\ref{fig:krtr}, the filled triangles show the fraction of samples whose $|n_{c1}|$ is smaller than $|n_{i}|$, which should correspond with the ratio of Type SS. For $|\omega_{xy0}|=0.001\,\tilde{\omega}$, all samples whose $|n_{c1}|$ is smaller than $|n_{i}|$ actually show Type SS behaviors and vice versa. On the other hand, for $|\omega_{xy0}|=0.1\,\tilde{\omega}$, there are some samples whose $|n_{c1}|$ is smaller than $|n_{i}|$ yet do not show Type SS behavior; for $n_i = -0.3 \,\tilde{\omega}$, there are only several Type SS samples out of 8000 samples, which cannot be seen in Fig.~\ref{fig:krtr}(a), and for $|n_i| \gtrsim 0.4\,\tilde{\omega}$, the fractions of Type SS for $|\omega_{xy0}|=0.1\,\tilde{\omega}$ are smaller than those for $|\omega_{xy0}|=0.001\,\tilde{\omega}$. This may be because $|\omega_{xy0}|=0.1\,\tilde{\omega}$ is not small perturbation, and the spin might have escaped from the basin of attractor of Type SS behavior.

Last we consider the Type SW samples. The time when the spin starts to
wobble roughly corresponds with $t_{rs}$ of Type R in
Fig.~\ref{fig:krtr-2}; the center of wobbling $n_{w}$ and its amplitude
vary from sample to sample.  As in the case of Type R, there are some
samples which start to wobble after several times of $t_{rGH-}$
where $K_{r} \ll 0.1$. Wobbling behaviors of such samples are similar to
those which start wobbling around $t_{rGH-}$.  We remark that there are
two qualitatively different Type SW behaviors. When $|n_i| \lesssim
0.4\,\tilde{\omega}$, the spin of Type SW sample oscillates almost
periodically. However, when $n_i = -0.5\,\tilde{\omega}$ and $K_{r} \ll
0.1$, we find some samples that show ``chaotic" oscillations as an
example shown in Fig~\ref{fig:krtr}(b-3).
%
%
\section{Discussion}
\label{sec:discussion} 

In the present work, we study the minimal model for the rattleback
dynamics, i.e., a spinning rigid body with a
no-slip contact ignoring any form of dissipation.
We have reduced the original dynamics to
the simplified dynamics (\ref{eq:diff-ne-1})--(\ref{eq:diff-ne-3})
with the three variables.  The assumptions and/or approximations used in
the derivation are (1) the amplitudes of the oscillations are small,
(2) the coupling between the spin and the oscillations does not depend
on the spin, and (3) the time scale for the spin change is much longer
than the oscillation periods.  It is interesting to note that the last
assumption is apparently analogous to that used in the derivation of an
adiabatic invariant for some systems under slow change of an external
parameter if the spin variable is regarded as a slow parameter.  
In the present case with this separation of time scales, the
dynamics conserves the ``Casimir invariant" $C_{I}$ of
Eq.~(\ref{eq:casimir}).  

Our simplified dynamics can be compared with some previous
works. Based on Bondi's formulation \cite{Bondi1986}, Case and Jalal
obtained the growth rates $\delta_{x}$ and $\delta_{y}$ of the pitching
and the rolling amplitudes around the $x$ and $y$ axes, respectively, at a
small constant spin and small skewness \cite{Case2014}. Their results
can be expressed as
\begin{equation}
\delta_{x} = \frac{n}{2}K_{p},\label{eq:case-instab1}\quad
\delta_{y} = -\frac{n}{2}K_{r},
\end{equation}
using our notations. The factor $1/2$ comes from the choice of the variables; they chose the contact point co-ordinates, while we choose the oscillation energies, which are second order quantities of their variables.

Moffatt and Tokieda \cite{MoffattTokieda2008} obtained equations for the
oscillation amplitudes of pitching and rolling, $P$ and $R$, and the
spinning $S$ for small spin and skewness as
\begin{equation}
\frac{d}{d\tau}
\begin{pmatrix} P \\ R \\ S\end{pmatrix} 
= \begin{pmatrix} R \\ \lambda P \\ 0\end{pmatrix}\times
\begin{pmatrix} P \\ R \\ S\end{pmatrix}
= \begin{pmatrix} \lambda PS \\ -RS \\ R^2 - \lambda P^2\end{pmatrix}
, \label{eq:toki-prs}
\end{equation}
where $\tau$ is rescaled time, and $\lambda$ is the squared ratio of the
pitch frequency to the roll frequency. Equation (\ref{eq:toki-prs}) is
equivalent with Eqs.~(\ref{eq:diff-ne-1})--(\ref{eq:diff-ne-3}); again
the difference comes from choice of the variables. The mathematical
structures of Eq.~(\ref{eq:toki-prs}) have been investigated recently in
more detail by Yoshida \textit{et al.} \cite{Yoshida2016} in
connection with the Casimir invariant and chaotic behavior of the
original dynamics.

\begin{figure}
\includegraphics[width=7.5cm]{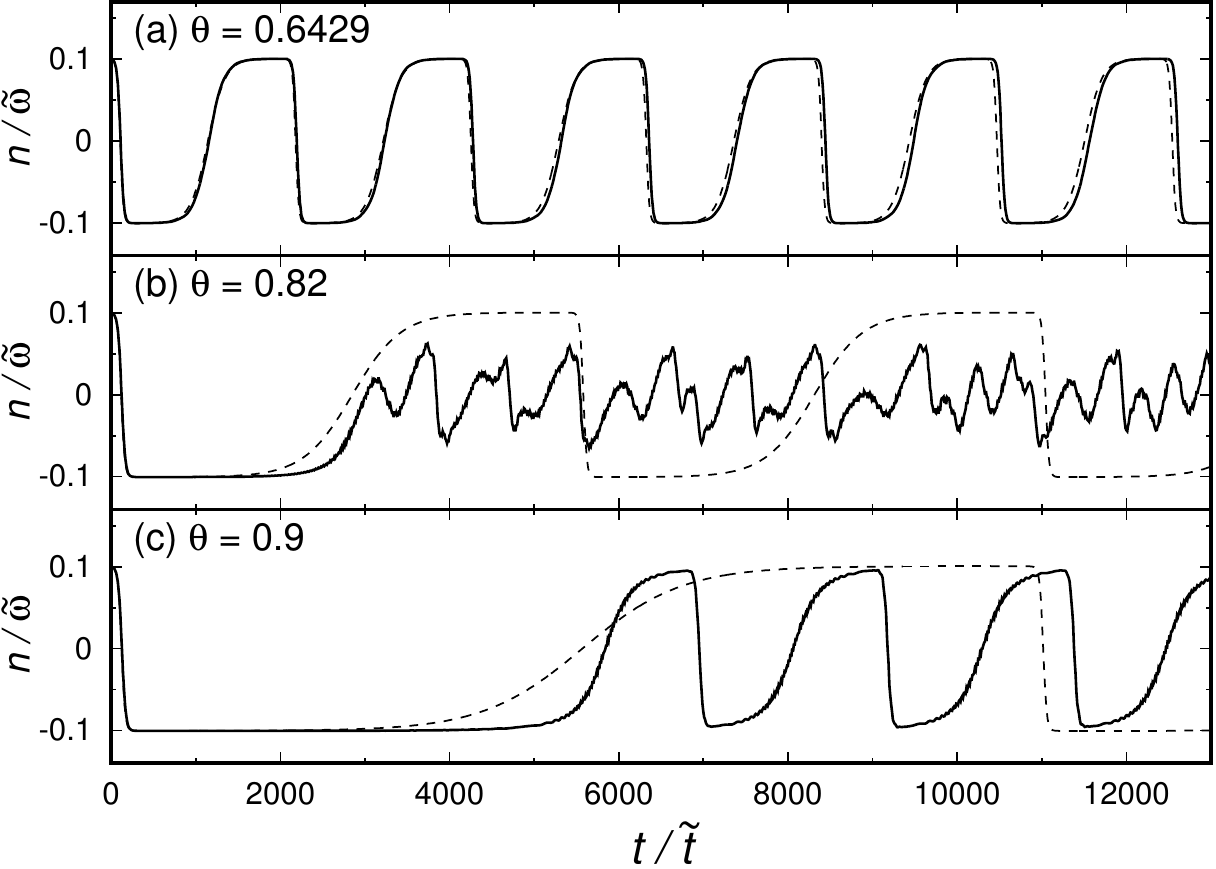} 
\caption{\label{fig:periodic} Three types of spin behaviors after the
first reversal period in the small spin regime with $n_{i} =
0.1\tilde{\omega}$, $|\omega_{xy0}|=0.01\tilde{\omega}$. (a) A
quasi-periodic behavior with the parameter set GH ($\theta = 0.6429$),
(b) a chaotic behavior with $\theta = 0.82$, (c) a quasi-periodic
behavior with a period shorter than the first one with $\theta =
0.9$. All the other parameters for (b) and (c) are the same as GH.
The dashed lines show the spin evolutions for the corresponding
simplified dynamics, where $\overline{E}_p(0) = Ma^2[\alpha (|\omega_{xy0}|\cos\psi_{p})^2 + \beta (|\omega_{xy0}|\sin\psi_{p})^2]/2$, $\overline{E}_r(0) = 3\times10^{-5}Ma^2\tilde{\omega}^2$, and $\overline{n}(0) = n_{i}$.}
\end{figure}
After the first round of spin reversals, our simplified
dynamics (\ref{eq:diff-ne-1})--(\ref{eq:diff-ne-3}) repeats
itself and shows periodic
behavior as well as the dynamics studied by Moffatt and Tokieda
Eq.~(\ref{eq:toki-prs}) because the system with only three variables has
two conservatives, i.e., the total energy and the Casimir
invariant.  However, the Casimir invariant is an approximate one in the
original dynamics, and invariant only under the approximations
given at the beginning of this section.  The Casimir ``invariant''
actually varies and the original system shows aperiodic behaviors.

A few examples for longer time evolutions of spin $n(t)$ are
given in Fig.~\ref{fig:periodic} for the system with the parameter set
GH except for the curvature in the rolling direction $\theta=0.6429$
 (a) for GH, $0.82$ (b), and $0.9$ (c) along with those by the
corresponding simplified dynamics.
The first example (a) almost shows a
periodic spin reversal behavior as is expected by the simplified
dynamics.  It is, however, only quasi-periodic
with  fluctuating periodicity.  
The second  example  (b) does not show a periodic behavior;
the initial spin reversal till $t/\tilde t\approx 100$ is nearly the
same with (a), but after the time of the second spin
reversal around $t/\tilde t\approx 3000$, it turns into chaotic,
deviating from the simplified dynamics.
The third example (c) may look similar to (a) but is peculiar;
it shows a quasi-periodic behavior after the initial round of
spin reversals, and its periodicity is
much shorter than that by the simplified dynamics.

The simplified dynamics seems to work reasonably well for the case of
smaller $\theta$ in (a) but fails for larger $\theta$ close to $1$ in
(b) and (c).
This indicates that the approximations or assumptions
used to derive the simplified dynamics are not valid for 
the larger curvature in the rolling direction $\theta$;
as the radius of curvature $1/\theta$ becomes small and close to 1, i.e.,
the height of the center of mass, the restoration force for the rolling
oscillation becomes weak.
This should result in the rolling oscillation with larger amplitude and
the slower frequency, thus the assumptions (1) and (3) given at the
beginning of this section may not be good enough.

The fact that the system shows a different behavior
after the first round of spin reversals is reminiscent of the existence of
attractors, which is normally prohibited in a conserving system by
Liouville theorem.  In the present system, however, the theorem is
invalidated by the non-holonomic constraint due to the no-slip condition
Eq.~(\ref{eq:no-slip}) \footnote{
The no-slip condition should be violated in the situations where the
ratio of the vertical and the inplane components of the contact force,
i.e., $F_{\parallel}\equiv\bm F\cdot\bm u$ and $F_{\perp}\equiv|\bm{F} -
(\bm{F}\cdot \bm{u})\bm u|$, exceeds the friction coefficient.  The
ratio $F_\perp/F_\parallel$ becomes large when the angular momentum
around $\bm u$ changes.  In the cases given in Fig.~\ref{fig:periodic},
its largest value is around 0.2. 
}.
As mentioned already, the existence of strange attractors in
an energy conserving system with a non-holonomic constraint has
been studied by Borizov et al. \cite{Borisov2014}, and chaotic behavior
in the rattleback system has been discussed in connection with the
Casimir invariant by Yoshida \textit{et al.} \cite{Yoshida2016}. 

%
\section{Summary and conclusion}
\label{sec:conclusion} 

We have performed the theoretical analysis and numerical
simulations on the minimal model of rattleback.  By reformulating Garcia
and Hubbard's theory \cite{GarciaHubbard1988}, we obtained the concise
expressions for the asymmetric torque coefficients, Eqs.~(\ref{eq:Kp})
and (\ref{eq:Kr}), gave the compact proof to the fact that the pitching
and the rolling generate the torques with the opposite sign, and reduced
the original dynamics to the three-variable dynamics by a physically
transparent procedure.

Our expressions for the asymmetric torque coefficients are
equivalent to those by Garcia and Hubbard, but we explicitly elucidate
that the ratio of the two coefficient for the pitching and the rolling
oscillation is proportional to the squared ratio of those frequencies.
Since the pitching frequency is significantly higher than that of the
rolling for a typical rattleback, the time for reversal to one spin
direction (or unsteady direction) is much shorter than that to the other
direction (or steady direction); the spin reversal for the latter
direction is not usually observed in a real rattleback due to dissipation.  

The simulations on the original dynamics for various parameter
sets demonstrate that Garcia-Hubbard formulas for
the first spin reversal time (\ref{eq:trgh-p}) and (\ref{eq:trgh-m}) are good in the case of
small initial spin and small oscillation for both the unsteady and the steady directions. The deviation from the formula is especially large for the steady direction in the fast initial spin and  small $K_r$ regime, where the
rattleback may not reverse and shows a variety of dynamics, that
includes steady spinning, periodic and chaotic wobbling.  

In conclusion, the rattleback is simple but shows fascinatingly rich
dynamics, and keeps attracting physicists' attention.

\bibliographystyle{apsrev}
%

\end{document}